\documentclass[twoside]{dis04}
\def\runauthor{M. Wing}
\def\shorttitle{Substructure dependence of jet cross sections}
\begin{document}

\title{SUBSTRUCTURE DEPENDENCE OF JET CROSS SECTIONS \\
AT HERA AND DETERMINATION OF $\alpha_s$
}

\author{MATTHEW WING \\
(On behalf of the ZEUS collaboration)}

\address{Bristol University, ZEUS, DESY\\
Notkestrasse 85, 22607 Hamburg, Germany\\
E-mail: wing@mail.desy.de }

\maketitle

\abstracts{The substructure of jets has been studied in terms of the jet shape and 
subjet multiplicity and these quantities used to tag gluon- and quark-initiated jets. 
Cross sections are presented for gluon- and quark-tagged jets which exhibit the 
expected behaviour of the underlying parton dynamics.
The value of $\alpha_s(M_Z)$ of \mbox{$\alpha_s(M_Z) = 0.1176 \pm 0.0009 ({\rm stat.}) 
^{+0.0009}_{-0.0026} ({\rm exp.}) ^{+0.0091}_{-0.0072} ({\rm th.})$} was extracted from 
the measurements of jet shapes in deep inelastic scattering.}

\section{Introduction} 

Through measurements of the jet substructure, knowledge of the final-, and hence, 
initial-state partons can be gleaned~\cite{substructure_paper}. This in turn allows 
for the enrichment of samples of events in a particular type of hard sub-process and 
direct probes of their dynamics, rather than probes of all sub-processes summed 
together. 

Jet substructure is here studied using two variables: the 
mean integrated jet shape, $<\psi(r)>$, defined as the averaged fraction of the jet 
transverse energy inside a cone $r$; and the subjet multiplicity, 
$<n_{\rm subjet}>(y_{\rm cut})$, defined as the average number of subjets contained 
in a jet at a given resolution scale, $y_{\rm cut}$. Using either or both of these 
variables, jets can be categorised as either gluon- or quark-like, as gluon jets 
are expected to be broader and have a higher multiplicity due to the larger 
colour factor. Similarly in events with two jets, both jets can be categorised. 
Knowing the identity of the two final-state partons then reduces the possible number of 
initial-state parton reactions which could have caused the observed event, thereby 
accessing specific sub-processes.

The substructure of jets is here studied along with the dependence of jet cross 
sections on the substructure and an extraction of $\alpha_s$.

\section{Results}

Events were selected in both the deep inelastic scattering (DIS) and photoproduction 
regimes which are characterised, respectively, by a large and small virtuality of the 
photon, $Q^2$. The photoproduction kinematic region is $Q^2 < 1$~GeV$^2$ and the 
photon-proton centre-of-mass energy between 142 and 293~GeV with a jet reconstructed 
in the region $E_T^{\rm jet} > $ 17 GeV and $-1 < \eta^{\rm jet} < 2.5$. Both 
inclusive-jet and dijet samples were considered, where the second jet has to have 
a transverse energy above 14 GeV. In DIS events, jets were also reconstructed in the 
region \mbox{$E_T^{\rm jet} > $ 17 GeV} and $-1 < \eta^{\rm jet} < 2.5$ in events with 
$Q^2 >$~125~GeV$^2$.

\begin{figure}[!thb]
\begin{center}
~\epsfig{file=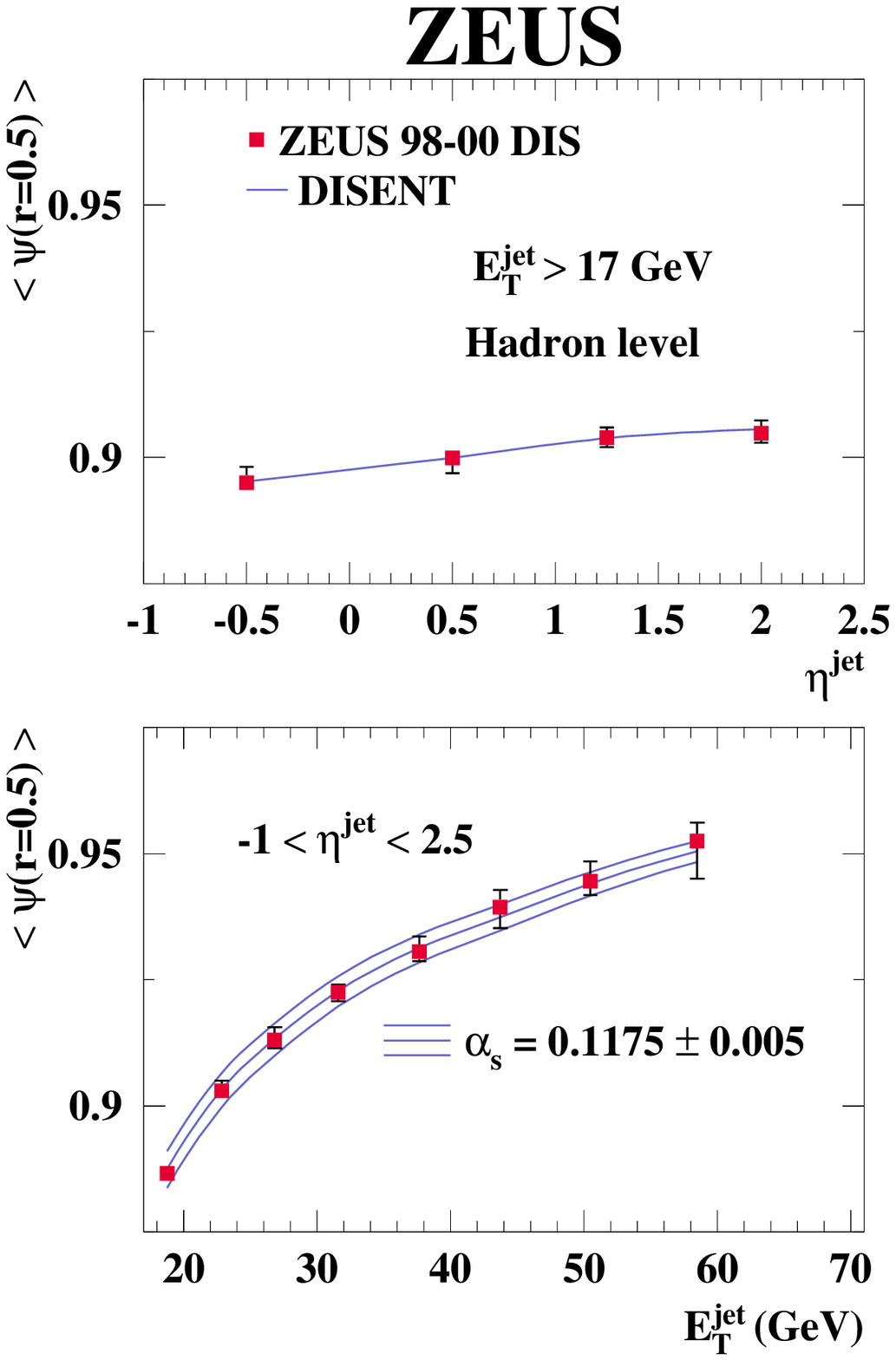,height=9cm}
~\epsfig{file=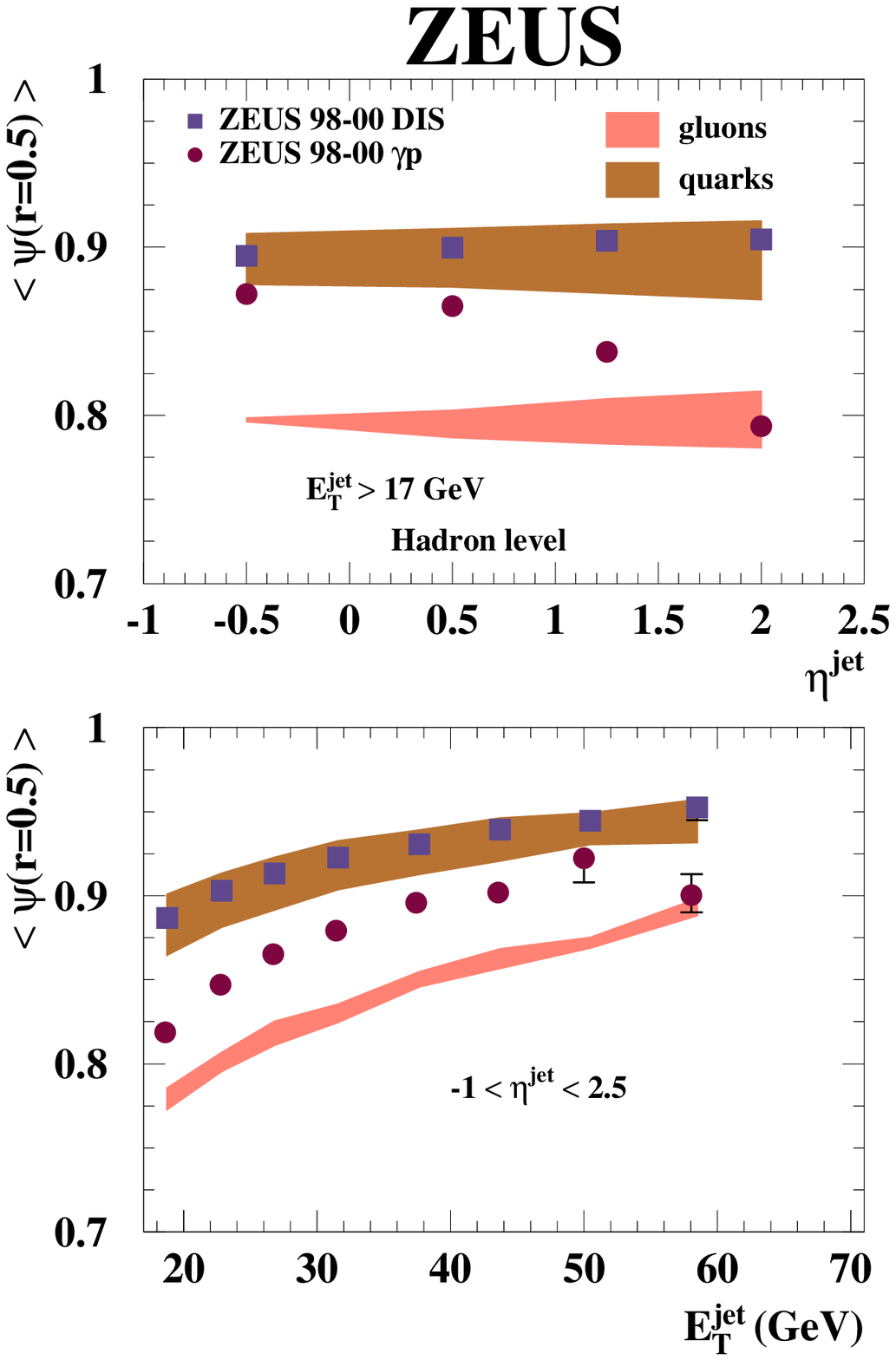,height=9cm}
\caption[*]{Measured mean integrated jet shape for a fixed value of $r=0.5$ for 
both DIS and photoproduction samples. The DIS measurements are compared with NLO 
QCD and the DIS and photoproduction samples are compared with MC predictions for 
the shapes of quark and gluon jets.
\label{fig:shapes}}
\end{center}
\end{figure}

The mean integrated jet shape has been measured~\cite{substructure_paper} in DIS 
differentially in $E_T^{\rm jet}$ and $\eta^{\rm jet}$ and compared with NLO QCD as 
implemented in the DISENT program~\cite{disent}. In Figure~\ref{fig:shapes} the DIS 
data is shown for a fixed value of $r=0.5$ compared with the predictions from 
NLO QCD. The data, as for the measurements differentially in $E_T^{\rm jet}$ and 
$\eta^{\rm jet}$ (not shown), are well described by NLO QCD. Little dependence of 
$<\psi(r=0.5)>$ 
as a function of $\eta^{\rm jet}$ is seen, whereas a significant dependence on 
$E_T^{\rm jet}$ is observed. This is compared with three different predictions of 
NLO QCD using different values of the strong coupling constant, $\alpha_s$, 
demonstrating that the data can be used to extract a value of $\alpha_s$. The value 
$\alpha_s(M_Z)$ of \mbox{$\alpha_s(M_Z) = 0.1176 \pm 0.0009 ({\rm stat.}) 
^{+0.0009}_{-0.0026} ({\rm exp.}) ^{+0.0091}_{-0.0072} ({\rm th.})$} was extracted. 
This value is consistent with other measurements at HERA and the world average. 
The uncertainty on the theoretical prediction is significantly larger than the 
experimental uncertainty of $^{+0.8}_{-2.2}\%$. 
Therefore with improved theoretical calculations, this measurement could become a 
very significant and precise determination of $\alpha_s$.

The DIS data is also shown in Figure~\ref{fig:shapes} compared with photoproduction 
data and Monte Carlo (MC) predictions~\cite{mc} for the shapes of quark and gluon 
jets. Compared with the flat dependence of $<\psi(r=0.5)>$ for the DIS data as a 
function of $\eta^{\rm jet}$, the photoproduction data decreases with increasing 
$\eta^{\rm jet}$. This is consistent with jets from quarks dominating in the rear 
region and jets from gluons dominating in the forward region in  $\eta^{\rm jet}$. 
Indeed at lowest and highest $\eta^{\rm jet}$, the data can be explained by only 
quark- and gluon-initiated jets, respectively. This is consistent with the increased 
fraction of resolved photon processes with increasing $\eta^{\rm jet}$. A similar 
variation can be seen as a function of $E_T^{\rm jet}$, where gluon-initiated jets 
(enriched in resolved photon processes) dominate at low $E_T^{\rm jet}$ and 
quark-initiated jets (enriched in direct photon processes) dominate at high 
$E_T^{\rm jet}$.

The differences in shapes quark- and gluon-initiated jets are used to enrich samples 
in the respective parton. A gluon-enriched sample (broad jets) is defined as those 
jets with $\psi(r=0.3) < 0.6$ and/or 
$n_{\rm subjet}(y_{\rm cut} = 5 \cdot 10^{-4}) \ge 6$. A quark-enriched sample 
(narrow jets) is defined as those jets with $\psi(r=0.3) > 0.8$ and/or 
$n_{\rm subjet}(y_{\rm cut} = 5 \cdot 10^{-4}) < 4$.

\begin{figure}[!thb]
\begin{center}
~\epsfig{file=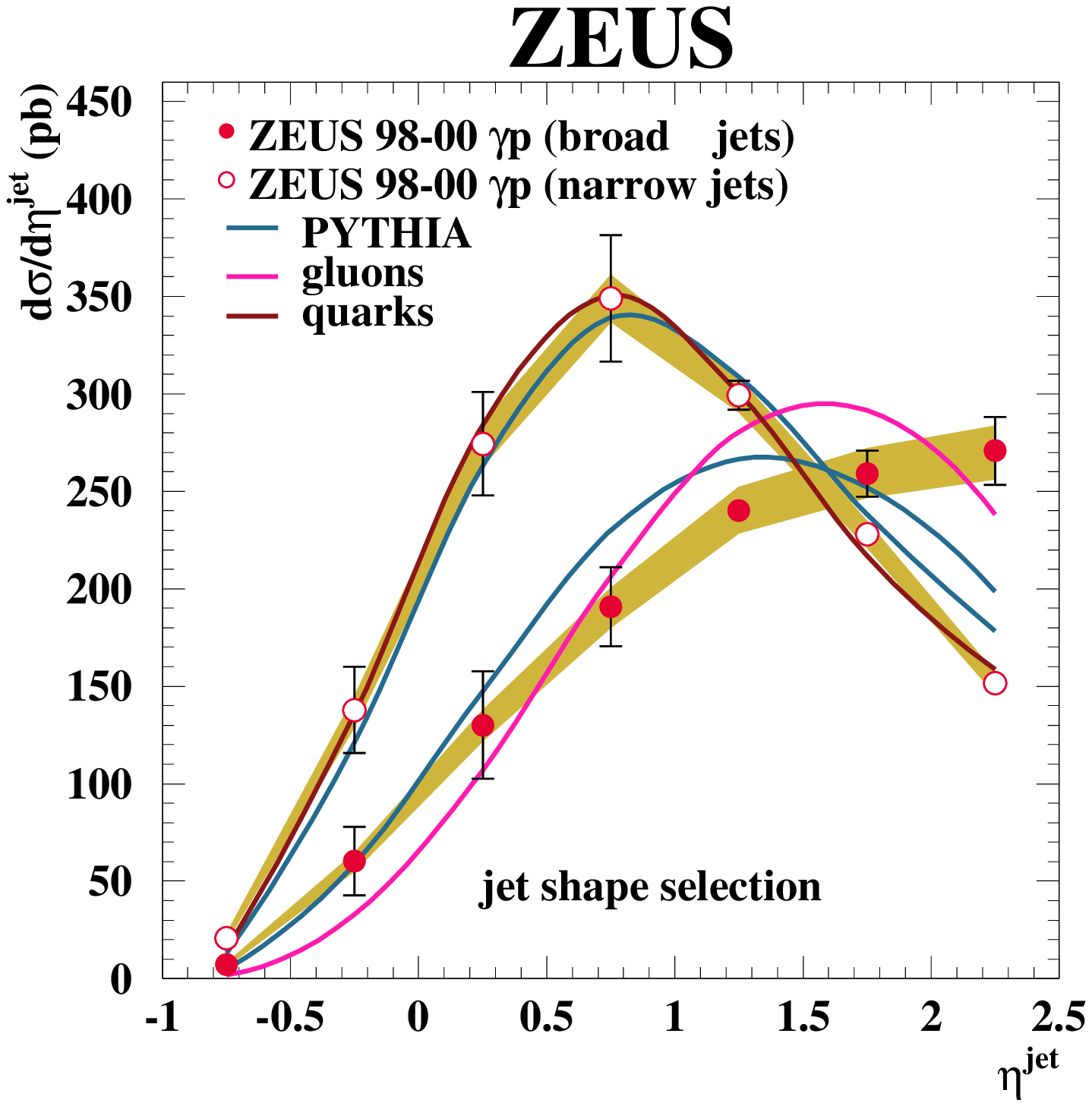,height=5.25cm}
~\epsfig{file=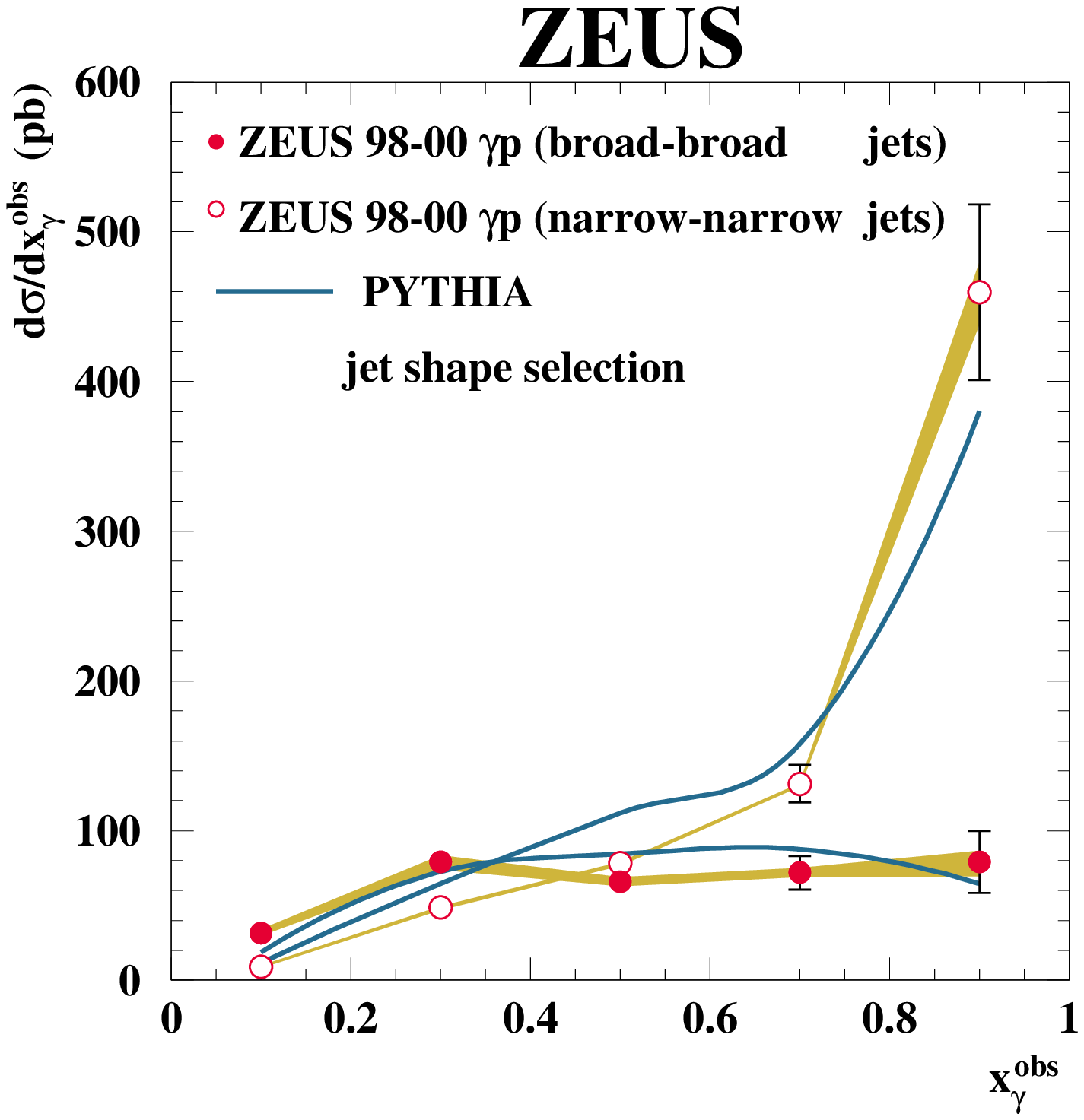,height=5.25cm}
~\epsfig{file=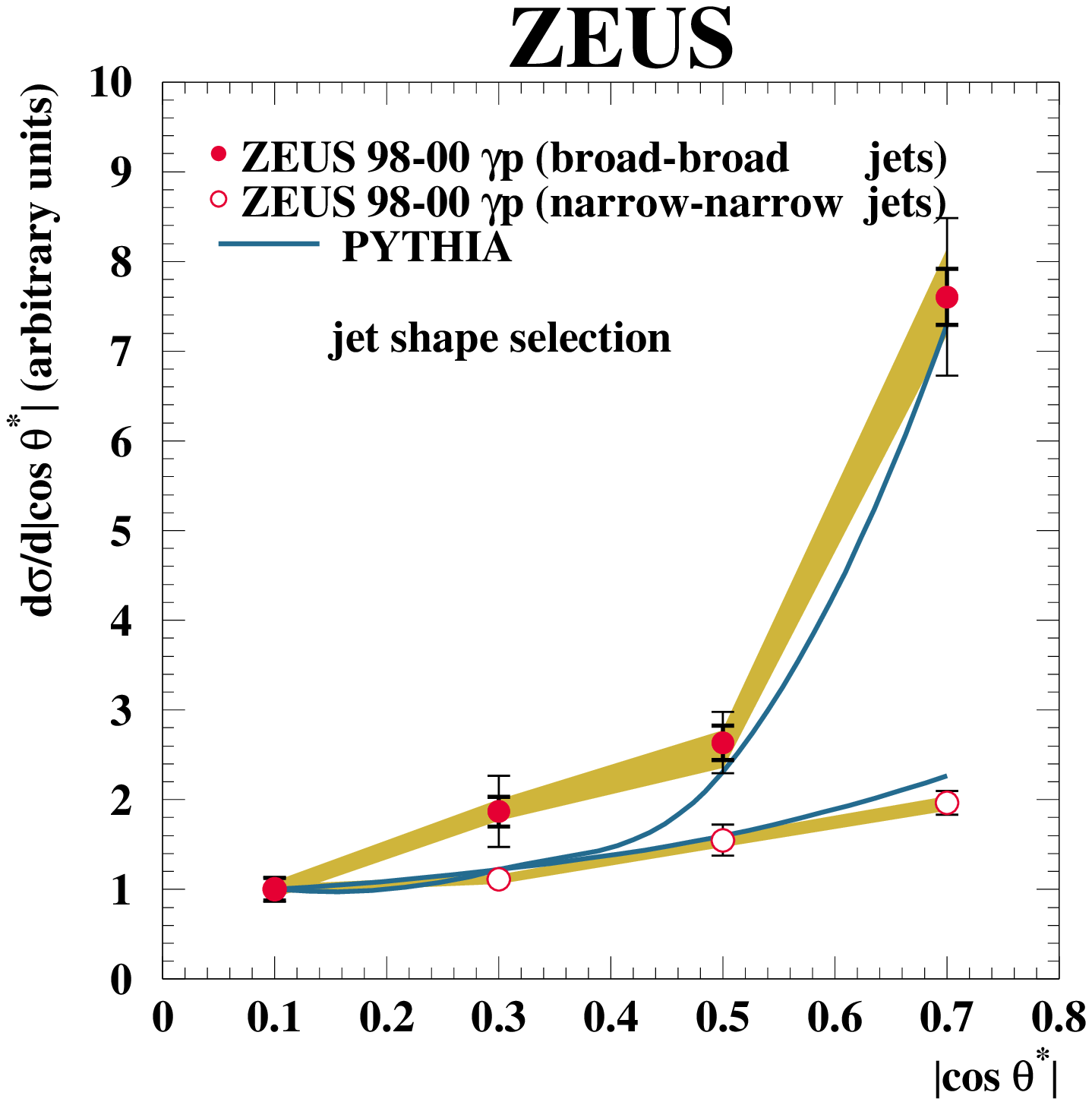,height=5.25cm}
~\epsfig{file=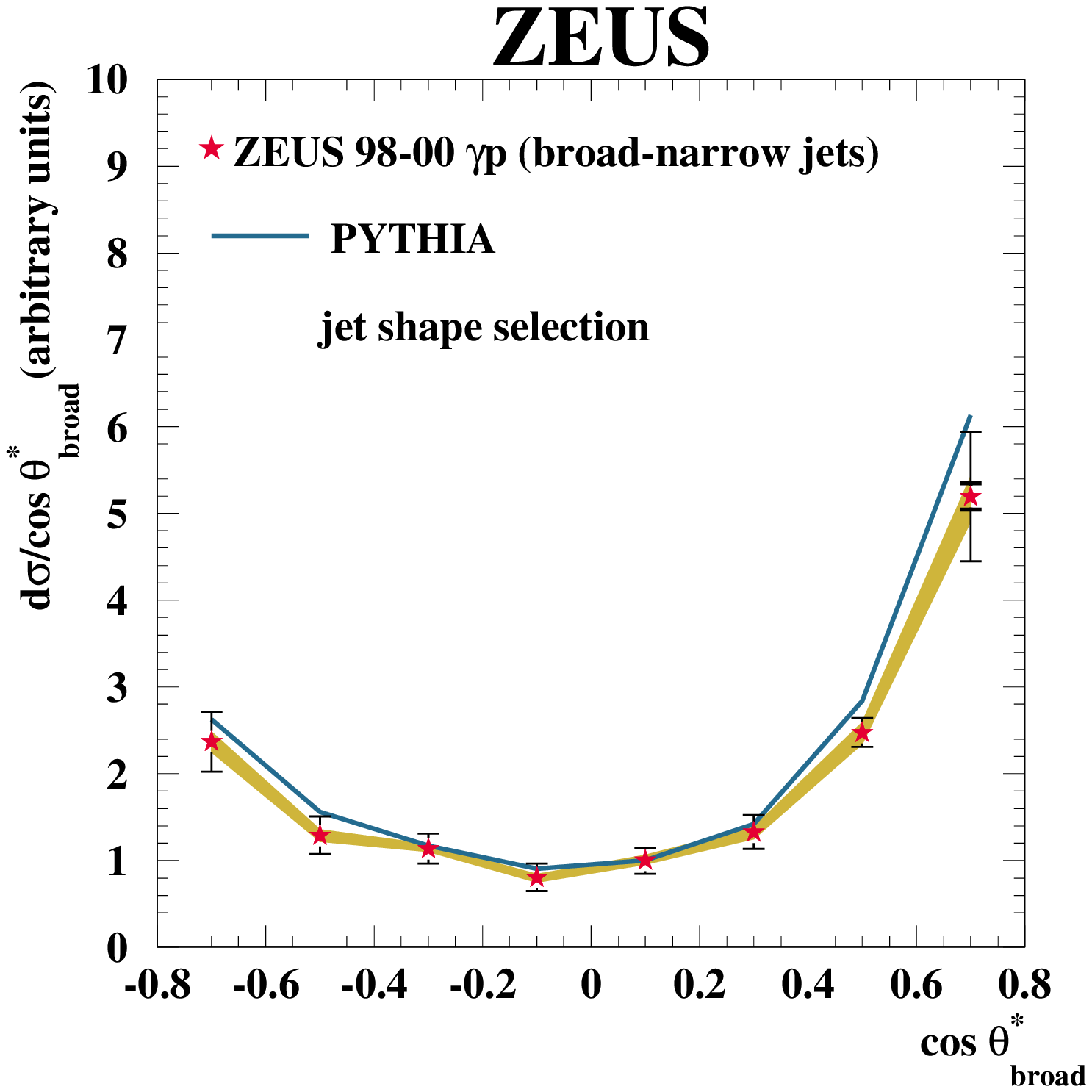,height=5.25cm}
\caption[*]{Cross-sections $d\sigma/d\eta^{\rm jet}$ for broad and narrow jets, 
$d\sigma/dx_\gamma^{\rm obs}$ and $d\sigma/d|\cos\theta^*|$ for two broad and two 
narrow jets and $d\sigma/d\cos\theta^*_{\rm broad}$ for one broad and one narrow jet.
\label{fig:xsecs}}
\end{center}
\end{figure}

The cross sections for broad and narrow jets in photoproduction are shown as a 
function of $\eta^{\rm jet}$ in Figure~\ref{fig:xsecs}. The distributions show a clear 
difference in shape with the cross section for narrow jets peaked at lower values 
of $\eta^{\rm jet}$. The data is reasonably well described by {\sc Pythia} which 
predicts 65\% of events with gluon-initiated jets in the broad sample and 96\% 
of events with quark-initiated jets in the narrow sample. 

By requiring two jets in the final state, purer samples of a certain sub-processes 
can be obtained and a richer collection of kinematic quantities can be measured to 
understand the parton dynamics. The cross section as a function of 
$x_\gamma^{\rm obs}$, the fraction of the photon's energy entering into producing 
the two highest $E_T$ jets, is shown in Figure~\ref{fig:xsecs} for two broad and 
two narrow jets. The distribution for two narrow jets in an event has a large cross 
section at high $x_\gamma^{\rm obs}$ indicative of direct photon events, demonstrating 
that the dominant sub-process is $\gamma g \to q \bar{q}$. For two broad jets in 
an event, the distribution is flat in $x_\gamma^{\rm obs}$ as would be expected from 
a large contribution from resolved photon events.

For two broad or two narrow jets in an event, only the absolute value of 
$\cos\theta^*$, where $\theta^*$ is the dijet scattering angle, can be determined as 
the jets are indistinguishable. The cross-section $d\sigma/d|\cos\theta^*|$ is shown 
in Figure~\ref{fig:xsecs} for $|\cos\theta^*| < 0.7$ and dijet invariant mass 
$M_{\rm jj} > 52$~GeV. The cross section for events with two narrow jets exhibits 
a mild rise with increasing $|\cos\theta^*|$ consistent with the exchange of a quark 
in the sub-process $\gamma g \to q \bar{q}$. The rapid rise for events with two broad 
jets indicates the exchange of a gluon as in the expected dominant process 
$q_\gamma g_p \to qg$. In events with one broad and one narrow jet, the sign of 
$\cos\theta^*$ can be measured, hence $d\sigma/d\cos\theta^*_{\rm broad}$ is shown 
in Figure~\ref{fig:xsecs}. The distribution exhibits an asymmetric distribution 
understood in terms of the dominant resolved photon sub-process $q_\gamma g_p \to qg$. 
As $\cos\theta^*_{\rm broad} \to +1$, the effect of $t$-channel gluon exchange can 
be seen whereas for $\cos\theta^*_{\rm broad} \to -1$, the effect of $u$-channel 
quark exchange is evident.

\section{Conclusion}

The substructure of jets in DIS and photoproduction at HERA has been measured. They 
are generally well described by QCD predictions. The measurement in DIS has been 
used to make an accurate and new extraction of the strong coupling constant. The 
substructure has also been used to categorise jets as quark- and gluon-initiated. 
This has allowed samples to be obtained which are enriched in a particular 
sub-process in both photoproduction and DIS; these data exhibit the expected 
behaviour of the underlying parton dynamics.


\begin{thebibliography}{0}

\bibitem{substructure_paper}
ZEUS Coll., S. Chekanov {\it et al.}, submitted to Nucl. Phys. DESY-04-072.

\bibitem{disent}
S. Catani and M.H. Seymour, Nucl. Phys. {\bf B485} (1997) 401;
S. Catani and M.H. Seymour, Nucl. Phys. {\bf B510} (1998) 503.

\bibitem{mc}
T. Sj\"{o}strand, Comp. Phys. Comm. {\bf 82} (1994) 74; 
Y. Azimov {\it et al.}, Phys. Lett. {\bf B165} (1985) 147;
G. Gustafson, Phys. Lett. {\bf B175} (1986) 453;
G. Gustafson and U. Pettersson, Nucl. Phys. {\bf B306} (1988) 746;
B. Andersson {\it et al.}, Z. Phys. {\bf C43} (1989) 625.

\end{thebibliography}
\end{document}